# Symmetry Properties of Anisotropic Superfluids

James V. Lindesay[*] and Harry L. Morrison
Department of Physics
University of California
Berkeley, California  94720

Abstract

A system of equations is developed for a fluid with non-abelian local gauge symmetry. Anisotropy is introduced by requiring that the symmetry breaking preserves a restricted local gauge symmetry about a given direction in the gauge parameter space. A set of equations is proposed for the $^3$He-A system. Topological quantization conditions are discussed.

## Introduction

Since the discovery of superfluidity, the concept of a superflow velocity field has been fundamental to the understanding of macroscopic ordering. In $^4$He the superfluid velocity can be introduced as a gauge field under some suitable local internal symmetry group[1]. A plausible theory for this premise has been developed [1,2]

In this paper we shall discuss the locally gauge symmetric fluid in which the full symmetry has been partially broken, leaving only a one-parameter subgroup of transformations as the symmetry group of the system. For the case of $^3$He-A, the local group of spin transformations will be the internal symmetry group (as a factor group), with the weak spin-orbital coupling to the (L = 1) orbital states or possibly the coupling to an external magnetic field breaking the symmetry. Without reference to the particular interaction which breaks the symmetry a set of equations which describe the behavior of the gauge field is proposed. A fuller treatment of the hydrodynamic equations is given in reference [2].

## Anisotropic Systems with Local Symmetries

It will be assumed that the state of the system under consideration is described by a local order parameter $\psi$, which transforms under a representation of the symmetry group. The generators of the group $G_a$ satisfy the algebra

---

[*] Present Address:  Computational Physics Laboratory, Howard University, Washington, DC   20059

$$[G_a, G_b] = c_{ab}{}^d G_d \quad \text{(summation)}.$$

(1)

The behavior of the order parameter under an infinitesimal transformation given by the group parameter $\delta\alpha^b$ is given by

$$\delta\psi(\vec{x}, t) = \delta\alpha^b G_b \psi(\vec{x}, t).$$

(2)

The thermodynamic parameters can be derived from a free energy f, which depends upon the order parameter and any additional thermodynamic variables which are invariant under the symmetry group:

$$f = f(\psi, \partial_\mu \psi, q_s).$$

(3)

The equations which govern the hydrodynamics of the system then follow from the stationarity requirement:

$$\delta \int f \, d^4 x = 0.$$

(4)

However, if the symmetry is local (gauge parameters have space-time dependence) the derivatives must be replaced by gauge-covariant derivatives,

$$D_\mu = \partial_\mu - W_\mu^a(\vec{x}, t; \underline{\alpha}) G_a.$$

(5)

In addition, the free energy may have dependence on the field strength defined by

$$F_{\mu\nu} = [D_\mu, D_\nu] = \left[-\partial_\mu W_\nu^a + \partial_\nu W_\mu^a + W_\mu^b c_{bd}{}^a W_\nu^d\right] G_a.$$

(6)

The general form of the free energy (suppressing the dependence on invariant parameters) will be

$$f = f(\psi, D_\mu \psi; F_{\mu\nu}).$$

(7)

It is convenient to define the conjugate parameter field

$$\Pi^\alpha \equiv \frac{\partial f}{\partial D_\alpha \psi}(\psi, D_\mu \psi; F_{\mu\nu}).$$

(8)

Then the gauge invariant momentum density is defined by

$$\vec{g} \equiv -\Pi^0 \vec{D}\psi = -\Pi^0 \vec{\nabla}\psi + \rho_a W^a$$
$$\rho_a \equiv \Pi^0 G_a \psi.$$

(9)

It should be noted that the tensorial character of the densities is rank 1 (vector), rather than the rank 2 (tensor) densities often introduced in the literature [3,4]. The gauge potentials transform inhomogeneously under the symmetry transformation

$$\delta W_\mu^b = \partial_\mu (\delta\alpha^b) + \delta\alpha^d c_{da}{}^b W_\mu^a.$$

(10)

The symmetry condition for a system which maintains invariance under the full group of transformations

$$\frac{\delta f}{\delta \alpha^r} = 0$$

(11)

has been previously examined[2]. If the symmetry is broken this condition will not hold in general. However, if a one parameter subgroup of symmetry transformations remains, parameterized by

$$\delta\alpha^b = \delta\theta\, n^b$$

(12)

then a partial symmetry of the system is maintained:

$$\frac{\delta f}{\delta \theta} = 0.$$

(13)

Here the direction vector *n* is the tangent vector of the one-parameter symmetry subgroup[5]. The effective generator of this subgroup becomes

$$\mathbf{G} = n^r G_r \equiv \aleph.$$

The symmetry condition Eq. (13) will imply a gauge current relation which will be explored shortly.

The group algebra admits a vector and scalar multiplication for its elements. If $\mathbf{V} = V^a G_a$ and $\mathbf{U} = U^a G_a$ are elements of the algebra, then

$$[\mathbf{V}, \mathbf{U}] = V^a c_{ab}{}^d U^b G_b$$

(14)

defines a vector product in the space. Similarly

$$trace(\mathbf{V}\mathbf{U}) \equiv (\mathbf{V}, \mathbf{U})$$

(15)

defines a symmetric scalar product. For elements of the adjoint representation $(G_a)_b{}^d = c_{ab}{}^d = (C_a)_b{}^d$, the scalar product defines a metric tensor:

$$g_{ab} = Tr\, \mathbf{C}_a \mathbf{C}_b = c_{ad}{}^f c_{bf}{}^d$$
$$(\mathbf{V}, \mathbf{U}) = V^a g_{ab} U^b \equiv V_b U^b = V^a U_a.$$

(16)

It is convenient to define the mixed tensor $\wp_{\perp a}{}^b(\mathbf{n})$ which projects into the subspace orthogonal to **n**:

$$\wp_{\perp a}{}^b(\mathbf{n}) \equiv \delta_a^b - \frac{n_a n^b}{(\mathbf{n}, \mathbf{n})}$$
$$\wp_{\perp a}{}^d \wp_{\perp d}{}^b = \wp_{\perp a}{}^b, \quad n^a \wp_{\perp a}{}^b = 0 = \wp_{\perp a}{}^b n_b.$$

(17)

The gauge potentials may be decomposed into longitudinal and transverse components with respect to the group space:

$$W_\mu^a = \left(\frac{W_\mu^b n_b}{(\mathbf{n},\mathbf{n})}\right) n^a + W_{\perp\mu}^a \equiv W_{\|\mu} n^a + W_{\perp\mu}^a.$$

(18)

From the invariance of the inner product, the condition $n_a \partial_\mu n^a = 0$ holds. Using this and the transformation, Eq. (10), the components of the gauge potential transform as follows:

$$\delta W_{\|\mu} = \partial_\mu (\delta\theta)$$
$$\delta W_{\perp\mu}^a = \delta\theta \left[\partial_\mu n^a + n^b c_{bd}{}^a W_{\perp\mu}^d\right]$$

(19)

The potential $W_{\|\mu}$ transforms as does the gauge vector potential for an abelian symmetry (eg. electromagnetism). The behavior under transformation of $W_{\perp\mu}^a$ is not simple rotation, due to the terms involving gradients of $\mathbf{n}$.

A set of equations will now be developed for the specialized case in which the structure constants are represented by the fully antisymmetric Levi-Cevita tensor:

$$c_{bd}{}^a = \varepsilon_{bd}{}^a$$
$$\varepsilon_{bd}{}^a = \begin{cases} 1 & for\ (b,d,a)\ cyclic\ permutations\ of\ (1,2,3) \\ -1 & for\ (b,d,a)\ odd\ permutations\ of\ (1,2,3) \\ 0 & otherwise \end{cases}$$

The metric is then represented $g_{ab} = -2\delta_{ab}$. For this case the direction vector $\mathbf{n}$ will be represented by $\ell$. The normalization condition for the direction vectors will be chosen to be

$$\sum_{a=1}^{3} \ell^a \ell^a = 1$$
$$(\ell,\ell) = -2 = \ell_a \ell^a.$$

(20)

Also, in correspondence with much of the literature on $^3$He-A an orthonormal triad of vectors will be introduced in the parameter space

$$(\phi_{(1)}, \phi_{(2)}, \ell) \quad , \quad \sum_a \phi^a_{(A)} \phi^a_{(B)} = \delta_{AB}$$

$$\ell = [\phi_{(1)}, \phi_{(2)}] \quad or \quad \ell^a = \phi^b_{(1)} \varepsilon_{bd}{}^a \phi^d_{(2)}.$$

(21)

Then the Eqs. (19) can be integrated to give

$$W_{\perp \mu}^{a}(\vec{x},t;\theta \ell) = \ell^b \varepsilon_{bd}{}^a \partial_\mu \ell^d + W_{(1)\mu}(\vec{x},t) \phi^a_{(1)}(\theta) + W_{(2)\mu}(\vec{x},t) \phi^a_{(2)}(\theta)$$

where

$$\phi^a_{(1)}(\theta) = \cos\theta\, \phi^a_{(1)} + \sin\theta\, \phi^a_{(2)}$$
$$\phi^a_{(2)}(\theta) = -\sin\theta\, \phi^a_{(1)} + \cos\theta\, \phi^a_{(2)}$$

(22)

Alternatively these equations can be written in an equivalent form. One defines the functions $W_{(+)\mu}{}^a$ by

$$W_{(+)\mu}^{a}(\vec{x},t) \equiv W_{(1)\mu}(\vec{x},t) \phi^a_{(1)} + W_{(2)\mu}(\vec{x},t) \phi^a_{(2)}.$$

(23)

Then

$$W_{\perp \mu}^{a}(\vec{x},t;\theta \ell) = \ell^b \varepsilon_{bd}{}^a \partial_\mu \ell^d + W_{(+)\mu}^{d}(\vec{x},t) \left[ \exp(\theta\, \ell^b\, \underline{\varepsilon}_b) \right]_d^a.$$

(24)

Either of these two forms demonstrates the decomposition of $W_\perp$ into a term invariant under the symmetry transformation (due to the $\partial_\mu \ell^d$ inhomogeneous term) and a term which rotates under the adjoint representation of the one parameter symmetry subgroup

$$W_{\perp \mu}^{a}(\vec{x},t;\theta \ell) = \ell^b \varepsilon_{bd}{}^a \partial_\mu \ell^d + W_{(R)\mu}^{a}(\vec{x},t;\theta).$$

(25)

The extra term $W^a_{(R)\mu}$, which does not appear in the literature for the perpendicular component of the fields $W^a$, is due to the gauge structure of the equations. The invariant term has a form similar to that which appears in the literature (see, for instance, reference 6).

In analogy to electromagnetism, the field strengths $F_{\mu\nu}$ will be decomposed as follows

$$F^a_{ij} \equiv -\sum_{k=1}^{3} \varepsilon_{ijk} \Omega^a_k$$

$$F^a_{0j} \equiv \mathcal{E}^a_j .$$

(26)

Additionally, the canonically conjugate field strengths will be defined

$$(\Lambda_a)_k \equiv \sum_{i,j=1}^{3} \frac{\partial f}{\partial F^a_{ij}} \varepsilon_{ijk}$$

$$(\kappa_a)_k \equiv 2 \frac{\partial f}{\partial F^a_{0k}}$$

(27)

and the 4-vector gauge potential will be decomposed into a "scalar" potential and a "three vector" potential, $W^a_\mu = (-v^a, W^a)$. Following the notation in Eq. (18), with

$$V_{(G)} \equiv \frac{V^a \ell_a}{(\ell, \ell)} = -\frac{1}{2} \ell_a V^a .$$

the equations which relate the gauge potentials to the field strengths are given by

$$\vec{\Omega}_{(G)} = \vec{\nabla} \times \vec{W}_{(G)} + \frac{1}{2} \ell^a \varepsilon_{abd} \vec{\nabla} \ell^b \times \vec{\nabla} \ell^d - \vec{W}_{(1)} \times \vec{W}_{(2)}$$

$$\vec{\mathcal{E}}_{(G)} = -\frac{1}{C_o} \frac{\partial}{\partial t} \vec{W}_{(G)} - \vec{\nabla} v_{(G)} + \ell^a \varepsilon_{abd} \vec{\nabla} \ell^b \left( \frac{1}{C_o} \frac{\partial \ell^d}{\partial t} \right) - v_{(1)} \vec{W}_{(2)} + v_{(2)} \vec{W}_{(1)} .$$

(28)

These equations are the analogue of the equation for $\nabla \times v_{(G)}$ in Mermin and Ho[7] and the equation for $\partial v_{(G)}/\partial t$ in Cross[8]. It should be noted that the field strengths $\Omega_{(G)}$ and $\mathcal{E}_{(G)}$ are gauge invariant under the restricted set of transformations of Eq. (12). The velocities $C_o$ are put in for dimensional convenience and can be set to unity. The behavior of the fields $\Omega, \mathcal{E}, \Lambda, \kappa$ is determined by the equations of motion and boundary conditions.

The partial symmetry condition $\delta f/\delta \theta = 0$ defines a constraint relation for gauge currents given by

$$\mathcal{J}^\mu_a = \Pi G_a \psi = (\rho_a, \vec{\mathcal{J}}_a).$$

(29)

Then the supercurrents satisfy the equations

$$\partial_\mu \mathcal{J}^\mu_{(G)} = -\frac{1}{2} \ell^a \varepsilon_{ab}{}^d \left[ W^b_{(R)\mu} \mathcal{J}^\mu_d + \vec{\Omega}^b \cdot \vec{\Lambda}_d - \vec{\mathcal{E}}^b \cdot \vec{\kappa}_d \right]$$

(30)

Thus due to the non-abelian character of the full group, the gauge currents are not in general conserved quantities.

## Quantization Equations

The topology of the manifold of the group transformations which maps into the coordinate space representation of the system determines many of the characteristic properties without regard to the specifics of the underlying dynamics. Among these properties are certain discrete quantum numbers which characterize the degree of the mapping of the gauge parameter space upon the configuration of the system. A set of quantum numbers analogous to "vortex number" or "monopole moment" will be developed which will characterize the structure of various stable topologies of the system.

For the 3-parameter space under consideration, there are two simple closed manifolds which can be directly explored. These simple compacts manifolds which can be imbedded in a 3-space are closed curves (one-dimensional manifolds) and closed simply-connected surfaces (two-dimensional manifolds). More complicated structures (for instance, surfaces with "handles" or multiple connectivity) will not be examined at present, but can similarly be explored.

To begin, topological structures of one dimension will be examined. For the restricted class of transformations, a differential parametric transformation can be defined:

$$\ell_b \, d\alpha^b = d\theta \, \ell_b \ell^b.$$

(31)

If $c_a$ is a (single loop) closed curve in the compact parameter space coordinatized by $\alpha$, then this implies (for periodicity $2\pi$) that

$$\oint_{c_\alpha} \ell_b \, d\alpha^b = 2\pi \ell_b \ell^b.$$

(32)

For the mapping $\alpha=\alpha(x,t)$, a (single loop) closed curve $c_x$ in x-space does not necessarily map singly into the curve $c_a$. However, the following does hold for locally integrable coordinations:

$$\oint_{c_\alpha} \ell_b \vec{\nabla}\alpha^b \cdot d\vec{x} = \oint_{N_{(1)}c_\alpha} \ell_b d\alpha^b = 2\pi N_{(1)} \ell_b \ell^b.$$

(33)

Here $N_{(1)}$ is the degree of the mapping[9], and represents an integer which characterizes the topology of the mapping of the configurational structures on the group parameter space. This relation will be shortly expressed in terms of the gauge vector potential.

If $\alpha$ and $\beta$ are elements in the group multiplication space of a Lie transformation group, there exists a multiplication rule which results in another element of this space given by

$$(\beta \cdot \alpha)^b = \phi^b(\beta;\alpha).$$

(34)

The function $\phi$ is at least doubly differentiable. It is convenient to define a nonsingular matrix $\Theta_a^b$ given by

$$\Theta_a^b(\alpha) \equiv \frac{\partial}{\partial \beta^a}\phi^b(\beta;\alpha)\bigg|_{\beta=\mathbf{e}}$$

(35)

where $\mathbf{e}$ is the identity element coordinatized as the origin of the group parameter space. Then, as shown in [2], the gradients of the gauge parameters are related to the gauge potentials:

$$\begin{aligned}\vec{\nabla}\alpha^b &= \vec{A}^a(\alpha)\Theta_a^b(\alpha) \\ \vec{W}^b(\vec{x},t;\alpha) &= \vec{A}^b(\alpha) + \vec{B}^b(\vec{x},t;\alpha) \\ \vec{W}^b(\vec{x},t;\mathbf{e}) &= \vec{B}^b(\vec{x},t;\mathbf{e})\end{aligned}$$

(36)

The functions $B^b$ transform homogeneously under the group of transformations (as an element of the algebra)

$$\delta\vec{B}^b(\vec{x},t;\alpha) = \delta\alpha^a c_{ad}{}^b \vec{B}^d(\vec{x},t;\alpha),$$

(37)

whereas the functions $A^b$ transform inhomogeneously (as gauge potentials)

$$\delta \vec{A}^b(\alpha) = \vec{\nabla}(\delta\alpha^b) + \delta\alpha^a c_{ad}{}^b \vec{A}^d(\alpha).$$

(38)

It can be shown[5] that a coordinatization of the group parameters of the one-parameter subgroup can be found such that $\alpha^b(\theta) = \theta\, \ell^b$. For these coordinates, the matrix $\Theta_a{}^b = \Theta^{(R)}{}_a{}^b$ simplifies

$$\Theta_a^{(R)b}(\theta, \ell) = \delta_a^b.$$

(39)

In addition, the relation (37) implies that $\delta B_{(G)}(x,t;\theta\,\ell) = 0$. Thus from relations (36) and (39).

$$\vec{A}_{(G)}(\theta) = \vec{W}_{(G)}(\vec{x},t;\theta) - \vec{B}_{(G)}(\vec{x},t;0) = \vec{W}_{(G)}(\vec{x},t;\theta) - \vec{W}_{(G)}(\vec{x},t;0)$$
$$\ell_b \vec{\nabla}\alpha^b = \vec{A}_{(G)}(\theta)\ell_b \ell^b = \left[\vec{W}_{(G)}(\vec{x},t;\theta) - \vec{W}_{(G)}(\vec{x},t;0)\right]\ell_b \ell^b.$$

(40)

Therefore, combining Eqs. (33) and (40) one obtains the following conditions

$$\oint_{C_x} \left[\vec{W}_{(G)}(\vec{x},t;\theta) - \vec{W}_{(G)}(\vec{x},t;0)\right] \cdot d\vec{x} = 2\pi N_{(1)}.$$

(41)

This relation is the analogue of the quantization of circulation for the abelian $^4$He system.

In addition to closed curves, a 3-parameter space admits closed surfaces. Thus topological structures occur which are not possible in a 1-parameter space. If the space is parameterized such that the integrated solid angle is $4\pi$, then

$$\oiint_{\Sigma_\alpha} \frac{1}{2} \ell^b \varepsilon_{ad}{}^b d\ell^a d\ell^d = 4\pi \ell_b \ell^b,$$

(42)

where $\Sigma_\alpha$ represents a closed 2-dimensional surface in the group parameter space. Similar to the closed curves previously mentioned, for the mapping $\alpha = \alpha(x,t)$, a closed surface $\Sigma_x$ need not singly cover the closed surface $\Sigma_\alpha$, but must be mapped an integral number of times through the diffeomorphism

$$\oiint_{\Sigma_\alpha} \frac{1}{2} \ell^b \varepsilon_{ad}{}^b \frac{1}{2} \frac{\partial(\ell^a, \ell^d)}{\partial(x^j, x^k)} dx^j dx^k = \oiint_{N_{(2)}\Sigma_\alpha} \frac{1}{2} \ell^b \varepsilon_{ad}{}^b d\ell^a d\ell^d = 4\pi N_{(2)} \ell_b \ell^b.$$

(43)

Here, $N_{(2)}$ is the degree of the mapping of $\Sigma_x$ onto $\Sigma_\alpha$, and $\partial(\ell^a, \ell^d)/\partial(x^j, x^k)$ is the Jacobian of the variable transformation. It is convenient to define the oriented differential element of area by

$$dA_i = \frac{1}{2} \varepsilon_{ijk} dx^j dx^k.$$

(44)

With this definition, the integrand of Eq. (43) can be written more compactly using

$$\frac{\partial(\ell^a, \ell^d)}{\partial(x^j, x^k)} dx^j dx^k = \left(\partial_j \ell^a \partial_k \ell^d - \partial_k \ell^a \partial_j \ell^d\right) dx^j dx^k = 2\left(\vec{\nabla}\ell^a \times \vec{\nabla}\ell^d\right) \cdot d\vec{A}.$$

(45)

The relation (43) can be expressed

$$\oiint_{\Sigma_x} \frac{1}{2} \ell_b \varepsilon_{ad}{}^b \left(\vec{\nabla}\ell^a \times \vec{\nabla}\ell^d\right) \cdot d\vec{A} = 4\pi N_{(2)} \ell_b \ell^b.$$

(46)

For the vectors normalized as in Eq. (20), this can be expressed as

$$\oiint_{\Sigma_x} \frac{1}{2} \ell_b \varepsilon_{ad}{}^b \left(\vec{\nabla}\ell^a \times \vec{\nabla}\ell^d\right) \cdot d\vec{A} = 4\pi N_{(2)}.$$

(47)

The equations developed can be directed related to the gauge invariant field strengths developed previously. From Eq. (28), the following quantization condition can be obtained:

$$\oiint_{\Sigma_x} \left[\vec{\Omega}_{(G)} + \vec{W}_{(1)} \times \vec{W}_{(2)}\right] \cdot d\vec{A} = 4\pi N_{(2)} = \iiint_{V_\Sigma} \vec{\nabla} \cdot \left[\vec{\Omega}_{(G)} + \vec{W}_{(1)} \times \vec{W}_{(2)}\right] d^3x.$$

(48)

Here $V_\Sigma$ is the volume enclosed by the surface $\Sigma_x$, and Gauss' theorem has been used in the last step. This equation expresses the possibility of "monopole like" structures in the field strengths $\Omega_{(G)}$. This relation is analogous to the quantization condition derived by Blaha in reference [10]. The additional term is due to the gauge group behavior of the system of equations.

Thus, there appears to be at least two types of simple topological structures in the non-abelian 3-parameter system under study. The quantum number $N_{(1)}$ represents a "vortex number" and the number

$N_{(2)}$ represents a "monopole moment" for the circulation field $\Omega_{(G)}$. The number $N_{(2)}$ does not occur in a system based on a one parameter group of symmetry. The arguments presented do not preclude the possibilities of more complicated topologies but are due entirely to the gauge structure of the theory.

## Conclusion

The equations developed demonstrate the richness of the gauge theoretical approach to systems with internal symmetries. The form of many of the hydrodynamic equations in the superfluid $^3$He physics can be obtained simply using these methods. In addition, a direct method for extracting topological quantization relations is obtained. These equations provide a relatively model independent approach to anisotropic superfluids.

## Acknowledgements

One of us (J.V.L.) wishes to acknowledge the Chancellors Distinguished Post-Doctoral Fellowship of the University of California, Berkeley.

Author's note:

This manuscript was completed by the authors in 1983. However, the submission process was not completed due to the relocation of one of the authors (J.V.L.). The manuscript is here being submitted in its original form in memory of my mentor and colleague, Prof. Harry Morrison, who recently passed.